\documentstyle[12pt,/usr/local/lib/tex/inputs/divers/epic/epic,
/usr/local/lib/tex/inputs/divers/eepic/eepic]{article}
\begin{document}

\textheight20cm
\textwidth14cm
\baselineskip0.8cm
\begin{titlepage}

\begin{center}
{\normalsize \bf Centre de Physique Th\'{e}orique\footnote{Unit\'e
Propre
de Recherche 7061}\\
          CNRS-Luminy-case 907-CPT\\
       F-13288 Marseille Cedex 9-France}\\[5mm]
{\large \bf  CRITICAL FLUCTUATION OPERATORS FOR A QUANTUM MODEL OF
FERROELECTRIC}\\[5mm]

{ \normalsize \bf Antony Car\footnote{Ecole Nationale Superieure de
Physique de Marseille,\hspace*{1cm}\\
 Domaine Universitaire de Saint-J\'{e}rome,
         13397 Marseille Cedex 30}
 and Valentin A. Zagrebnov\footnote{ and D\'{e}partement de Physique,
Universit\'{e} d'Aix-Marseille {I\hskip -0.5pt I}\\ \hspace{1cm}
         E-mail address: zagrebnov@cpt.univ-mrs.fr}}
                                        \vspace*{5mm}\\
\end{center}

\begin{abstract}

We consider dependence of the critical fluctuation operators on the
rate of interaction decay for an exactly soluble model of the
quantum anharmonic crystal of ferroelectric type. The critical
exponents for the abnormal behavior of the displacement and momentum
fluctuation operators on the critical line are calculated explicitly.

\vspace{10mm}\end{abstract}


\noindent Number of figures: 3

\noindent May 1994\\
CPT-94/P.3034

\noindent anonymous ftp or gopher: cpt.univ-mrs.fr

\end{titlepage}

\section{ \bf Introduction}

\hspace{0.6cm}
The quantum nature of the structural phase transitions of certain type
of ferroelectric models was a subject of some recent experimental [1]
and theorical [2--4] papers.

The first of these quantum effects is connected with the long debate
''wash out''of the phase transition by the quantum fluctuations for
the case of the light atoms [2--4]. The second effect, discovered in
[2], is displayed by the ''squeezing''of the fluctuation operator
conjucated to the one with abnormal fluctuations.

The calculations in [2,3] are guided by the idea of the fluctuation
operators algebras developed in [5--7]. In the frame of this approach,
one can control the fluctuation properties of the quantum systems in
the critical region as well as outside of it. As it was shown in
[2,3], the fluctuation operators generated by displacements and
momenta of atoms build up abelian algebra
of normal fluctuations above the critical line, while on the critical
line the fluctuation operators can be abnormal and the algebra can be
non-abelian.

The abnormal fluctuation operators are characterized by the exponents
showing a deviation from the standart square root inherent to normal
fluctuation operators. As far as these exponents manifest the
long-range correlations in the critical region, it is important to
know their dependence on the dimension of the lattice { \it Z\hskip
-4pt Z \hskip 0.2pt}$^{d}$
and on the range of the interaction between atoms [8].

The aim of the present paper is to elucidate the behavior of the
fluctuation operators on the critical line for the model [2,3] when
the harmonic force matrix elements $\phi_{ll'}$ have a long-range
inverse power-law decay:\\ $\displaystyle \phi_{ll'} \simeq
|l-l'|^{-(d+\sigma)}\ \ {\rm for}\ \ |l-l'|\rightarrow\infty$.\\
Scrutinizing the quantum central limit [5--7] for the basic
fluctuation
operators (displacement and momentum), we show that for the case of
$0<\sigma<2$ \linebreak the corresponding exponents are modified by
$\sigma$, while for $\sigma\geq2$ \linebreak the fluctuation operators
belong to
the same class universality as those for\linebreak the short-range
interactions:
$\phi_{ll'}=0 {\ \rm for\ }|l-l'|>R \ {\rm or\ }\phi_{ll'}\simeq
\exp{(-|l-l'|/R)}$\linebreak for $|l-l'| \rightarrow \infty$.

The paper is organized as follows. In Section 2 we define the model
and recall its thermodynamic properties. Definition of the
fluctuation operators and their properties are collected in Section~3. In the
Sections~4 and~5 we formulate and demonstrate our main results about
properties of the critical fluctuation operators for
displacements and momenta on the critical line for the long-range
interaction case. We discuss our results in Section~6.
%
%
%
%
%
%
%
\newsavebox{\Bd}
\savebox{\Bd}{$ {\cal B}_{\rm d}=\{q:\  |q_{\alpha}| \leq \pi,\
\alpha=1, 2,..., d\}$}
\newsavebox{\idl}
\savebox{\idl}{$ I_{d}(c_{\Lambda})$}
\newsavebox{\idls}
\savebox{\idls}{$ I_{d}^{(>\epsilon)}(c_{\Lambda})$}
\newsavebox{\idlm}
\savebox{\idlm}{$ I_{d}^{(<\epsilon)}(c_{\Lambda})$}
\newsavebox{\ide}
\savebox{\ide}{$ I_{d}(c^{*})$}
\newsavebox{\idem}
\savebox{\idem}{$ I_{d}^{(<\epsilon)}(c^{*})$}
\newsavebox{\ides}
\savebox{\ides}{$ I_{d}^{(>\epsilon)}(c^{*})$}
\newsavebox{\bc}
\savebox{\bc}{$ \beta_{c}$}
\newsavebox{\depid}
\savebox{\depid}{$ (2\pi)^{d}$}
\newsavebox{\dqd}
\savebox{\dqd}{$ d^{d}q$}
\newsavebox{\ro}
\savebox{\ro}{$ \frac{1}{V}
               \frac{\lambda}{2\sqrt{\Delta(c)}}
                              \coth\frac{\beta\lambda}{2}
\sqrt{\Delta(c)}
              $}
\newsavebox{\rens}
\savebox{\rens}{\it I\hskip -2pt R}
\newsavebox{\zens}
\savebox{\zens}{ \it Z\hskip -3.8pt Z \hskip 0.2pt}

%
%
%
\newcommand{\fluc}[3]{\mbox{$
                      \frac{1}{V^{\frac{1}{2}+#1}}
                      \sum_{#2\in\Lambda}(#3_{#2}-\eta(#3))
                     $}}
\newsavebox{\del}
\savebox{\del}{$\Delta(c_{\Lambda})$}
\newsavebox{\cl}
\savebox{\cl}{$c_{\Lambda}$}
\newsavebox{\ce}
\savebox{\ce}{$c^{*}$}

\renewcommand{\u}[1]{\usebox{#1}}

	\section{ \bf Model and Phase Transition}
\hspace{0.6cm}
We consider an exactly soluble model invented in [9] to study
ferroelectrics showing displacement structural phase
transitions with general anharmonicity.\\
Let $\u{\zens}^{d}$ be d-dimensional lattice. Each lattice site $l\in
\u{\zens}^{d}$ associated with quantum particle with mass $m$,
position $Q_{l} \in \u{\rens}^{1}$ and momentum
$\displaystyle P_{l}=\frac{\hbar}{i}\frac{\partial}{\partial Q_{l}}$.
The local Hamiltonian $H_{\Lambda}$ for any finite subset $\Lambda
\subset \u{\zens}^{d}$ with $V=|\Lambda|$ is given by the operator
$$ H_{\Lambda}=
               \sum_{l\in\Lambda}\frac{P_{l}^{2}}{2m}
               + \frac{1}{4}\sum_{l, l'\in\Lambda}
                              \phi_{l-l'}\left(Q_{l}-Q_{l'}\right)^{2}
               + \frac{a}{2}\sum_{l\in\Lambda} Q_{l}^{2}
               + V W \left(
                           \frac{1}{V} \sum_{l\in\Lambda}Q_{l}^{2}
                     \right)
               - h \sum_{l\in\Lambda}Q_{l},  \vspace{-2mm}
$$
$$\eqno \rm(2.1)\vspace{-30mm}$$\\
acting on the Hilbert space
$\displaystyle {\cal H}_{\Lambda}=\bigotimes_{\Lambda}L^{2}
(\u{\rens}^{1})$.

The two first terms represent the harmonic lattice, while the $a$-term
and
W-term correspond to the anharmonicity. To describe ferroelectric
structural transition, we have to take a local double-well potential
with generate minimum. For example, $\ a>0$ and the function
$\displaystyle W(x)=\frac{b}{2}\exp\left(-\eta x\right)$ with $b,\
\eta>0 $ and $b$
and $\eta$ sufficiently large to destabilize the $a$-term for small
displacements $\{Q_{l}\}_{l\in \Lambda}$ [9].\\
The model (2.1) results from the {\it ansatz}
$$\sum_{l\in\Lambda}W\left(Q_{l}^{2}\right)\rightarrow V W
\left(\frac{1}{V}
\sum_{l\in\Lambda}Q_{l}^{2}\right), $$
applied to the local anharmonic potential
$W\left(Q_{l}^{2}\right)$ [9]. This {\it ansatz} gives an exactly
soluble
Hamiltonian (2.1) corresponding to the well-known concept of {\it
self-consistent phonon}, see, e.g.[10].
The model (2.1) is soluble in the sense that for all temperatures
$T\geq0$ and $ h \in \u{\rens}^{1}$, one can calculate the free
energy
density and the thermal averages explicitly [2]. \\
To analyse the possible critical behavior of the system (2.1), we
use the
explicit formula (see [2]) for the free energy density given by
\begin{eqnarray}
f(T,\ h)= \lim_{\Lambda}\left\{  \frac{1}{\beta V}
                                  \sum_{l\in\Lambda^{*}}
                                   \ln[\:2\sinh(\frac{\beta \lambda
                                               \Omega_{q}(c_{\Lambda})}{2})\:]
                         \right.
                 &+&  W(c_{\Lambda})- c_{\Lambda} W'(c_{\Lambda})
                                                       \nonumber \\
[5mm]
                 &-&     \left.           \frac{h^{2}}{2\u{\del}}
                        \right\},
                                \nonumber
\end{eqnarray} \vspace*{-20mm}\nopagebreak
$$\eqno\rm (2.2) \vspace{-3mm}$$
where $c_{\Lambda}=c_{\Lambda}(T,\ h)$ is a solution of the
self-consistency equation
$$  c=\langle\,\frac{1}{V}\sum_{l\in\Lambda}Q_{l}^{2}
     \,\rangle_{H_{\Lambda}(c, h=0)}
               = \frac{h^{2}}{\Delta^{2}(c)}
                +\frac{1}{V}\sum_{q\in\Lambda^{*}}\frac{\lambda}
{2\Omega_{q}(c)}\coth \frac{\beta \lambda}{2}\Omega_{q}(c)
                \eqno \rm(2.3)
$$
with
$$ \Omega_{q}^{2}(c)=\Delta(c) + \omega_{q}^{2}\eqno\rm (2.4)$$
$$ \Delta(c)=a+2W'(c)                          \eqno\rm (2.5)$$
$$ \omega_{q}^{2}=\tilde{\phi}(0)-\tilde{\phi}(q).\eqno\rm (2.6)$$
Here $\{\Omega_{q}\}_{q\in\Lambda^{*}}$ are the frequencies of the
approximating harmonic Hamiltonian [2,9]. The set $\Lambda^{*}$ given
by
$$\Lambda^{*}=\left\{q: \displaystyle \frac{2\pi}{N}n^{\alpha};\
n^{\alpha}=0,\ \pm1,\
\pm2,...,\ \pm(\frac{N}{2}-1),\ \frac{N}{2};\ \alpha=1,\ 2,...,\
d \right\}$$
is the dual lattice to the hypercube
$\Lambda=(-\frac{N}{2},\ \frac{N}{2}]^{d} $ with periodic boundary
conditions.
The function $\displaystyle
\tilde{\phi}(q)=\sum_{l\in\Lambda}\phi_{l}e^{-i\,q \cdot l}$ is the
Fourier transform of the periodically extended
$\phi_{l-l'}$, $\ \beta=1/kT$ ($T$ is temperature) and finally
\vspace{-3mm}
       $$\ \lambda=\hbar/\sqrt{m}  \eqno{(2.7)} \vspace{-3mm}$$
\noindent is quantum parameter of the model. The function $\Delta(c)$
 represents a gap in the spectrum (2.4) of the
self-consistent phonons.
The study of solutions of the equation (2.3) in the thermodynamic
limit $\Lambda\rightarrow \u{\zens}^{d}$ yields the phase diagram,
Fig.1.
In this limit the equation (2.3) transforms into\\
$$  c=\rho(T,\ \lambda)+I_{d}(c,\ T,\ \lambda), \eqno\rm (2.8)$$
where
$$ \rho(T,\ \lambda)=\lim_{\Lambda}
                               \frac{1}{V}
                               \frac{\lambda}{2\sqrt{\u{\del}}}
                               \coth\frac{\beta\lambda}{2}
                               \sqrt{\u{\del}}
                               +\frac{h^{2}}{\Delta^{2}(c)}
                                            \eqno {(2.9)}  $$
and
$$ I_{d}(c,\ T,\ \lambda)=\frac{\lambda}{(2\pi)^{d}}
                        \int_{\cal B_{\rm d}}d^{d}q\,\frac{1}
{2\Omega_{q}(c)}
                                \coth\frac{\beta\lambda}{2}
\Omega_{q}(c).
                                       \eqno\rm(2.10)
$$\vspace{5mm}

Here \u{\Bd} is the first Brillouin zone. \\
Solutions of the equation (2.8) belong to the domain $D=[\,c^{*},
\infty\ )$ where (see(2.3)--(2.6))
$$
c^{*}=\inf\{c: c\geq0,\  \Delta(c)\geq0\}.\eqno{(2.11)}
$$
Therefore, necessary condition that the system (2.1) for $h=0$
manifests phase transition is $\u{\ce}>0$, such that
$\Delta(\u{\ce}>0)=0$.\\ Let $ W:
\u{\rens}^{1}_{+}\rightarrow \u{\rens}^{1}_{+}$ be a monotonous
decreasing function with \linebreak $W''(c)\geq w \geq0$. Then this
condition is equivalent to the requirement of a double-well form of
the potential: $ a + 2W'(c)<0$,
destabilizing the $a$-term for $0 \leq c< \u{\ce}$. This means that
$\u{\ce}$ in (2.11) is solution of the equation $ a + 2W'(\u{\ce})=0$.
\\
Let $h=0$. Then for a fixed $\u{\ce}>0$ the line of the critical
temperature
(phase diagram) is defined by the condition $\rho(T_{c}(\lambda),\
\lambda)=\nolinebreak0$ or (see(2.8))
$$
c^{*}=I_{d}(c^{*},\ T,\ \lambda). \eqno {(2.12)}
$$
This critical line $T_{c}(\lambda)$ (or $\lambda_{c}(T)$) separates
the
phase diagram (see Fig. 1) into two open domains (I) and
({I\hskip -0.8pt
I}):
$$
          \left\{
                 \begin{array}{ll}
                    \rho(T,\ \lambda)=0 & {\rm if\ }(T,\ \lambda)\in
({\rm I})\ \cup\ \partial\ ({\rm I\hskip -0.8pt I})\\[4mm]
                    \rho(T,\ \lambda)> 0 & {\rm if\ }(T,\ \lambda)\in
({\rm I\hskip -0.8pt I}).
                 \end{array}
           \right.
 \eqno {(2.13)}$$\\
For fixed {\it T}, varying $\lambda$ across $\lambda_{c}(T),\ $the
state goes from \linebreak a regime~$\rho(T,\ \lambda)>0$ to a
regime~$\rho(T,\ \lambda)=0$
and the same occurs for fixed $\lambda$, varying {\it T}
across $T_{c}(\lambda)$. So, the phase transition is driven by
both the parameters $\lambda$ and $T$. Remark that for
$ T_{c}(\lambda_{c})=0$ (quantum limit), phase transition is only due
to quantum tunneling effect. Furthermore, only for the quantum
paramater $\lambda=\hbar/\sqrt{m} \rightarrow 0 $, the critical
temperature has classical value $T_{c}$ (classical limit).\\
In order to characterize the phases (I) and ({I\hskip -0.8pt I}),
 remark
that the Hamiltonian (2.1) for $h=0$ has the $\u{\zens}^{2}$ symmetry:
$\ Q_{l}\rightarrow - Q_{l}$, i.e,
$$
\eta_{\Lambda}(Q_{l}) \equiv \langle Q_{l}\rangle_{H_{\Lambda}(h=0)}=0.
$$
For $h\neq0$ the solution $\u{\cl}(T,\ h)>\u{\ce}$, see (2.3).
Therefore, equation (2.8) gets the form
$$
c=\frac{h^{2}}{\Delta^{2}(c)}+ I_{d}(c,\ T,\ \lambda). \eqno{(2.14)}
$$
Hence, the solution of (2.14) $c_{h}(T,\ \lambda)>\u{\ce}$. The
expectation value of the average displacement
$$
\eta(Q_{l}) \equiv \lim_{\Lambda}\langle Q_{l}\rangle_{H_{\Lambda}(h)}
           =\frac{h}{\Delta(c_{h})}\ \  .\eqno{(2.15)}
$$
Therefore, for the phase (I) $\cup\ \partial\ ({\rm I\hskip
-0.8pt I})$,
i.e, for the case $c^{*}\leq I_{d}(c^{*},\ T,\ \lambda)$ (see Fig.1)
we have $\displaystyle \lim_{h\to 0} c_{h}(T,\ \lambda)=c(T,\
\lambda)\geq\u{\ce}$. Hence,
$$
\lim_{h\to 0} \eta(Q_{l})=0. \eqno{(2.16)}
$$
For the phase ({I\hskip -0.8pt I}), i.e, for the case $c^{*}>
I_{d}(c^{*},\ T,\ \lambda)$ (see Fig.1) we have $\displaystyle
\lim_{h\to 0}
c_{h}(T,\ \lambda)= \u{\ce}$ and
$$
\rho(T,\ \lambda)= c^{*}-I_{d}(c^{*},\ T,\
\lambda)=\lim_{h\to 0} \frac{h^{2}}{\Delta^{2}(c_{h})} >0.
\eqno{(2.17)}
$$
Hence, for the average displacement (order parameter) one gets
$$
\eta_{\pm}(Q)= \lim_{h \to \pm 0} \eta(Q_{l})=\pm
\sqrt{\rho(T,\ \lambda)}\neq 0 . \eqno{(2.18)}
$$

Therefore, in the domain ({I\hskip -0.8pt I}) the model (2.1)
describes
both:\\
(i). the softening of the phonon mode: $\Omega_{q=0}(\u{\ce})=0$;\\
(ii). the spontaneous breaking of the $\u{\zens}^{2}$ symmetry
(displacement structural phase transition): $ \eta_{\pm}(Q)=\pm
\sqrt{\rho(T,\ \lambda)}\neq 0 $.
\section{\bf Algebra of Fluctuation Operators}
\hspace{0.6cm}
To investigate critical behavior of the model (2.1) on the line
$T_{c}(\lambda)$ we examine the fluctuations corresponding to the
algebra generated by the canonical operators $Q_{l}$ and $P_{l}$
with $l,\
l'\in \u{\zens}^{d}$.\\
A recent theory [5--7] introduces the notion of the algebra of
fluctuation operators, which is complementary to the well-known
algebra
of the observables at infinity [8].\\
Without going into details, for which we refer to [5--7], let us
present here main notions needed for the future calculations.

Consider $A_{i}=\tau_{i}(A)$ a copy of local operator $A$ translated
at the lattice point
$i\in \u{\zens}^{d}$ by the automorphism $\tau_{i}$. The fluctuation
operator $F_{\delta}(A)$ in the ergodic translation invariant
state
$\eta=\lim_{\Lambda}\eta_{\Lambda}$ on the algebra $\cal A$ of
quasi-local observables is unbounded operator on some
Hilbert space, given by
          $$ F_{\delta}=\lim_{\Lambda} \frac{1}{V^{\frac{1}{2}+\delta}}
                      \sum_{i\in\Lambda}(A_{i}-\eta_{\Lambda}(A_{i})),
                                                       \eqno \rm (3.1)
          \vspace{6mm}$$
where $\delta$ is defined by the clustering properties of the state
$\eta$ [5,6]. The limit (3.1) is in the sense of the central limit
Theorem in the state $\eta$ [5]. The existence of the fluctuation
operator
is technically guaranteed if one proves that the variance of the
fluctuations is not trivial:

         $$ 0<\ \lim_{\Lambda} \eta_{\Lambda} \left(
                      \left\{
                             \frac{1}{V^{\frac{1}{2}+\delta}}
                                      \sum_{i\in\Lambda}
(A_{i}-\eta_{\Lambda}(A_{i}))
                      \right\}^{2}
                                               \right)
             \ <\infty.      \eqno \rm (3.2)
          $$
The fluctuations critical exponent $\delta$ measures the deviation
from the standart central limit. It measures the degree of {\it
criticality} of the fluctuation operator.\vspace{3mm}\\
If $\delta>0$ then $F_{\delta}$ corresponds to {\it abnormal
critical fluctuations}.\vspace{-2mm}\\
If $\delta=0$ then $F_{\delta}$ corresponds to {\it normal
fluctuations}.\vspace{-2mm}\\
If $\delta<0$ then $F_{\delta}$ corresponds to {\it
supernormal (squeezed) critical fluctuations}.\vspace{4mm}

To characterize algebra of fluctuation operators one has to calculate
their commutators. Let $A$ and $B$ be two local operators, i.e, $A, B
\in \cal A$. The commutator of
fluctuation operator $F_{\delta}(A)$ and $F_{\delta'}(B)$ is defined
formally by \vspace{2mm} \\
\noindent
$\displaystyle
\lim_{\Lambda} \left[\frac{1}{V^{\frac{1}{2}+\delta}}\sum_{i\in\Lambda}
(A_{i}-\eta_{\Lambda}(A_{i}))
  ,\     \frac{1}{V^{\frac{1}{2}+\delta'}}\sum_{j\in\Lambda}(B_{j}-
\eta_{\Lambda}(B_{j}))
   \right]
$
\begin{eqnarray}
                          &=&    \lim_{\Lambda}\frac{1}{V^{1+\delta+
\delta'}}\sum_{i\in\Lambda}\left[A_{i},\ B_{i}\right]
                                                        \nonumber\\
[7mm]
                          &=&    \left\{
            \begin{array}{ll}
                        \eta\left([A, B]\right)  & {\rm if}\ \
\delta+\delta'= 0\\
                                      0          & {\rm if}\ \
\delta+\delta'>0 .\\
                                      undefined  & {\rm if}\ \
\delta+\delta'< 0
              \end{array} \nonumber
                                  \right.
\end{eqnarray} \vspace{-20mm}
$$
\eqno{(3.3)}
$$\\
Here we used local commutativity $\left[A_{i},\
B_{j}\right]=\delta_{ij}\left[A_{i},\ B_{i}\right] $ and the
weak-limit $\Lambda \rightarrow \u{\zens}^{d}$ in the state $\eta$.
Hence, for the ergodic state $\eta$ the equation (3.3) gives a
nontrivial canonical commutation relation
between the fluctuation operators $F_{\delta}(A)$ and $F_{\delta}(B)$
if $\eta([A, B])\neq0$. It indicates the quantum nature of the
fluctuations [2,3].

For our model (2.1) we consider two fondamental fluctuation operators
corresponding to the local displacement $Q_{l}$ and momentum $P_{l}$
operators of particles on the lattice $\u{\zens}^{d}$.
They are given by:
$$
F_{\delta}(Q)=\lim_{\Lambda}\frac{1}{V^{\frac{1}{2}+\delta}}
                              \sum_{l\in\Lambda}(Q_{l}-\eta_{\Lambda}
(Q_{l}))
\eqno \rm (3.4)
$$\vspace{-6mm}\\
and
$$
F_{\delta'}(P)=\lim_{\Lambda}\frac{1}{V^{\frac{1}{2}+\delta'}}
\sum_{l\in\Lambda}(P_{l}-\eta_{\Lambda}(P_{l})).
\eqno \rm (3.5)
$$\\
For the different regions of the phase diagram, we have to find
$\delta$ and $\delta'$ such that the variances in the sense of (3.2)
be nontrivial. The explicit calculation of the variances yields
$$
\lim_{\Lambda}\eta_{\Lambda}
                  \left(
                      \left\{
                             \frac{1}{V^{\frac{1}{2}+\delta}}
                               \sum_{l\in\Lambda}(Q_{l}-\eta_{\Lambda}
(Q_{l}))
                      \right\}^{2}
                  \right)
           = \lim_{\Lambda}\frac{1}{V^{2\delta}}
\frac{\lambda}{2\sqrt{\Delta(c_{\Lambda})}}
\coth\frac{\beta\lambda}{2}
\sqrt{\Delta(c_{\Lambda})}
\eqno\rm (3.6)
$$\vspace{-8mm}\\
and
$$
\lim_{\Lambda}\eta_{\Lambda}
                  \left(
                      \left\{
v                             \frac{1}{V^{\frac{1}{2}+\delta'}}
                               \sum_{l\in\Lambda}(P_{l}-\eta_{\Lambda}
(P_{l}))
                      \right\}^{2}
                  \right)
           = \lim_{\Lambda}\frac{1}{V^{2\delta'}}
                               \frac{\lambda m\sqrt{\Delta(c_{\Lambda}
)}}{2}
                               \coth\frac{\beta\lambda}{2}
\sqrt{\Delta(c_{\Lambda})},
\eqno\rm (3.7)
$$\\
where $c_{\Lambda}$ is a solution of the self consistent equation
(2.3). Then above the critical line $T_{c}(\lambda),\ $ where
$\lim_{\Lambda}\u{\del}>0$, the values $\delta=\delta'=0$ and both
momentum and displacement fluctuations are normal.\\
Below the critical line $T_{c}(\lambda),\ $ momentum fluctuation
operator $F_{\delta'=0}(P)$ is normal while displacement one
$F_{\delta}(Q)$ is abnormal with the critical exponent $\delta$
depended on the boundary conditions introduced by the external field
\linebreak $h_{\Lambda}={\hat h}/V^{\alpha},\ \alpha \geq0$, see [2,3].\\
For the short-range interactions $\phi_{l-l'}\ $ calculations on
the critical line $T_{c}(\lambda)$ show nontrivial exponents $\delta$
and $\delta'$ [2].

In the present paper, we restrict our study to the fluctuations on the
critical line for a quantum model of ferroelectic (2.1) with the
long-range interactions. It turns out that this type of
interaction can modify exponents $\delta$ and $\delta'$, (3.4), (3.5),
measuring fluctuation criticality.\\
To distinguish long-range and short-range interactions,
we consider \linebreak inverse power-law harmonic force matrix
$\phi_{l-l'}\ $,
decaying as $|l-l'|^{-(d+\sigma)}$ \linebreak for $|l-l'|\rightarrow
  \infty $.
For $0<\sigma<2$ exponent $\sigma$ characterizes long-range
interactions.
In this case, the Fourier transform (2.6) is given for $q\rightarrow0$
 by
$$
\tilde\phi(0)-\tilde\phi(q)=a^{\sigma}q^{\sigma}\ +o(q^{\sigma}),
\eqno \rm(3.8)
$$
\noindent where $a \equiv a(d,\ \sigma)>0$ and $ q=|q|$.
While for the values $\sigma\geq 2$ one gets for $\tilde\phi(q)$,
$q\rightarrow 0$, the same behavior as for
the short-range interactions which
includes potentials decaying exponentially and potentials with a
strictly finite-range, i.e, with a finite support: the Fourier
transform (2.6) is independent of the exponent $\sigma$ and is
determined for $q\rightarrow0$ by
$$
\tilde\phi(0)-\tilde\phi(q)=s^{2}q^{2}\ +o(q^{2}),\eqno \rm(3.9)
$$
\noindent where $ s \equiv a(d,\ \sigma\geq2)>0 $.
\section{\bf Temperature critical fluctuations}

\hspace{6mm}
On the critical line $\displaystyle \lim_{\Lambda}c_{\Lambda}
(T_{c},\
h=0)= c^{*}=I_{d}(c^{*}, T_{c}(\lambda), \lambda)$, i.e,
 $\rho(T_{c}(\lambda), \lambda)=0$, see (2.9).
Therefore, the asymptotic behavior of the gap $\Delta(c_{\Lambda}
(T_{c},\
h=0))$ is defined by the condition:
$$
       \lim_{\Lambda}\frac{1}{V}
                          \frac{\lambda}{2\sqrt{\Delta(c_{\Lambda})}}
                               \coth\frac{\beta_{c}(\lambda)
\lambda}{2}
                                          \sqrt{\Delta(c_{\Lambda})}
         = 0 .                 \eqno\rm (4.1)
$$\\
We distinguish two different parts of the critical region:
$T_{c}(\lambda)>0$ and $T_{c}(\lambda_{c})=0$, see Fig.1.\\
In the first case, the condition (4.1) is equivalent to
$$
\lim_{\Lambda}\frac{1}{V \beta_{c}(\lambda) \Delta(c_{\Lambda})}
= 0.
\eqno \rm (4.2)
\vspace{-4mm}$$\\
In the second case, (4.1) is equivalent to
$$
\lim_{\Lambda}\frac{\lambda}{2 V \sqrt{\Delta(c_{\Lambda})}}=0.
\eqno \rm (4.3)
\vspace{-4mm}$$\\
In the both cases, for $V \rightarrow \infty$ the gap
$\Delta(c_{\Lambda}(T_{c}(\lambda),\ h=0)) \simeq V^{-\gamma}$ with
 $0<\gamma<1$. The exponent $\gamma$ defines the values $\delta$ and
$\delta'$ in (3.6) and (3.7). In this section we consider
$T_{c}(\lambda)>0$.
\proclaim Lemma 4.1.
Let ($T, \lambda$) belongs to the critical line ($T_{c}(\lambda),
\lambda$) and \linebreak $T_{c}(\lambda)>0$. Then for $\ 0<\sigma<2$
the asymptotic volume behavior of the gap
$\Delta(c_{\Lambda}(T_{c}(\lambda),\ h=0))$ is defined by
$$
\gamma=    \left\{
                  \begin{array}{lc}
  \displaystyle \frac{\sigma}{d}   & \displaystyle \sigma<d<2
\sigma\\[7mm]
  \displaystyle \frac{\sigma}{d}+0 & \displaystyle   d=2\sigma  .  \\
[7mm]
  \displaystyle \frac{1}{2}        & \displaystyle   2\sigma<d
\end{array}
\right.
$$\par \vspace{10mm}
\noindent
{\it Proof.} First we represent the self consistent equation (2.3)
for
$h=0$ as:
\begin{eqnarray}
  \{c_{\Lambda}-c^{*} \} &+&  \{c^{*}-I_{d}(c_{\Lambda},\ T_{c}
(\lambda),\
\lambda) \} \nonumber \\[7mm]
                     &  + &  \{ I_{d}(c_{\Lambda},\ T_{c}(\lambda),\
\lambda)
                                    -\frac{1}{V}\sum_{q\neq0}
\frac{\lambda}{2\Omega_{q}(c_{\Lambda})}\coth \frac{\beta_{c}
(\lambda)
\lambda}{2}\Omega_{q}(c_{\Lambda})
                               \}
                                 \nonumber \\[7mm]
                     &  = &  \frac{1}{V}
                                       \frac{\lambda}{2\sqrt{\Delta
(c_{\Lambda})}}  \coth\frac{\beta_{c}(\lambda)\lambda}{2}
                                          \sqrt{\Delta(c_{\Lambda})}.
                                                    \nonumber
                  \end{eqnarray}\vspace{-21mm}
$$\eqno{(4.4)} \vspace{7mm}$$\\
Now we have to examine each term of the equation (4.4) to find the
 asymptotic of $\Delta(c_{\Lambda})$.
As far as $\Delta(c)\in C^{1}[0,\ \infty)$ and $W''(c) \geq w >0$,
see
(2.5), we can use the Taylor expansion to represent the first term as
$$
(c_{\Lambda}-c^{*})=\frac{\Delta(c_{\Lambda})}{2W''(\tilde{c}_{\Lambda})}.
\eqno \rm (4.5)
\vspace{-5mm}$$\\
Here $\tilde{c}_{\Lambda} \in (\u{\ce},\ \u{\cl})$.
The second term is $\u{\ce}-\u{\idl}=\u{\ide}-\u{\idl}$. Hence, for
small $ \epsilon>0$ and $\u{\del} \rightarrow 0$ we can
represent it as follows:
\begin{eqnarray}
\u{\ide}-\u{\idl}& = &  \u{\idem}-\u{\idlm}
                       +\u{\ides}-\u{\idls} \nonumber\\[7mm]
           & = & \frac{\u{\del}}{\u{\bc}\u{\depid}}
                 \int_{|q|<\epsilon}\u{\dqd}
                   \frac{1}{a^{\sigma}q^{\sigma}(\u{\del}+a^{\sigma}
q^{\sigma})} \nonumber\\[7mm]
           & - & (\u{\cl}-\u{\ce})\partial_{c}I_{d}^{(>\epsilon)}
(\hat{c}_{\Lambda}),\nonumber
\end{eqnarray}
where $\hat{c}_{\Lambda}\in (\u{\ce},\ \u{\cl})$.
Here
\begin{eqnarray}
\u{\idem}-\u{\idlm} &  = & \frac{\u{\del}}{\u{\bc}\u{\depid}}
                            \int_{|q|<\epsilon}\u{\dqd}
                           \frac{1}{a^{\sigma}q^{\sigma}(\u{\del}+
a^{\sigma}q^{\sigma})} \nonumber \\[7mm]
                    &  = & \frac{S_{d}\u{\del}}{\u{\bc}\u{\depid}}
                            \int_{0}^{\epsilon}dq\frac{q^{d-1}}
{a^{\sigma}q^{\sigma}(\u{\del}+a^{\sigma}q^{\sigma})},\nonumber
\end{eqnarray}\vspace*{-20mm}\nopagebreak
$$ \eqno{(4.6)}$$\\
where $S_{d}$ is the surface of the unit sphere in $\u{\rens}^{d}$.
Now we have to distinguish two cases:\\
(i). $2\sigma<d$. Then
$\displaystyle
                           \int_{0}^{\epsilon}dq\frac{q^{d-1}}
{a^{\sigma}q^{\sigma}(\u{\del}+a^{\sigma}q^{\sigma})}
                           = K_{\epsilon} \mbox{\ is \ finite}$.
Therefore,we get
$$
\u{\idem}-\u{\idlm} = \frac{S_{d}\u{\del}K_{\epsilon}}{\u{\bc}
\u{\depid}}.
$$\\
(ii). $\sigma<d \leq 2\sigma$. Consider first
$2\sigma=d$. In this case, the asymptotic of
the integral (4.6) has the form:
$$
\u{\idem}-\u{\idlm} =  \frac{2S_{d}}{\u{\bc}\u{\depid}a^{d}d}
                       \u{\del}
                       \ln
                \frac{\u{\del}+(a\epsilon)^{d/2}}{\u{\del}}.
      \vspace{-3mm}$$\\
For the case $\sigma<d< 2\sigma$ after the change of variable
$\displaystyle
x=\frac{a^{\sigma}q^{\sigma}}{\Delta}
$, we obtain
$$
\u{\idem}-\u{\idlm}  =
\frac{S_{d} \Delta^{d/\sigma-1}}{\u{\bc}\u{\depid}a^{d}\sigma}
\int_{0}^{ \frac{\scriptstyle
(a\epsilon)^{\sigma}}{\Delta}}\:\frac{x^{d/\sigma-2}}{1+x}\:dx.
                                        \eqno\rm(4.7)
\vspace{-3mm}$$\\
Hence, the asymptotic of (4.7) for $\u{\del} \rightarrow 0$ gets the
form
$$
\u{\idem}-\u{\idlm}= N_{1}(\u{\del})^{d/\sigma-1}+N_{2}\u{\del},
\eqno\rm(4.8)
\vspace{-7mm}$$\\
where $N_{1}=N_{1}(d,\ \sigma)$ and $N_{2}=N_{2}(d,\ \sigma)$.
To estimate the asymptotic of the last term in left-hand side of the
equation (4.4), we use again the representation
$I_{d}^{(<\epsilon)}+I_{d}^{(>\epsilon)}$.
It allows to calculate the difference between integral
$I_{d}$ and sum by the method proposed in [11]:\\
$\displaystyle
I_{d}(\u{\cl},\ ,T_{c}(\lambda),\ \lambda)
       -
\frac{1}{V}\sum_{q\neq0}\frac{\lambda}{2\Omega_{q}(\u{\cl})}\coth
\frac{\beta \lambda}{2}\Omega_{q}(\u{\cl})
$
$$
     = E(d,\ \sigma) V^{-(d-\sigma)/d} + o(V^{-(d-\sigma)/d}).
 \eqno\rm(4.9)
$$
Now we are in position to find the asymptotic volume behavior of
\u{\del}. It depends on the relation between parameters $\sigma$ and
$d$.\\
$a.\ \ \ 2\sigma<d.$ The equation (4.4) gets (for
$\u{\del}\rightarrow0$) the following form:\vspace{5mm}

$\displaystyle
\left\{ \frac{\Delta(c_{\Lambda})}{2W''(\tilde{c}_{\Lambda})}
        +\frac{S_{d}\u{\del}K_{\epsilon}}{\u{\bc}\u{\depid}}
\right.$
$$
\left.
        -\frac{\Delta(c_{\Lambda})}{2W''(\tilde{c}_{\Lambda})}
\partial_{c}I_{d}^{(>\epsilon)}(\hat{c}_{\Lambda})
        + E(d,\ \sigma) V^{-(d-\sigma)/d}
\right\}
        V\u{\bc}(\lambda)\u{\del}
        = 1.
\vspace{-3mm}$$\\
Therefore, we have
$$
                 \u{\del} \simeq V^{-1/2}.
\eqno\rm(4.10)
$$\\
$\displaystyle
       b.\ \ \ 2\sigma=d.
$
Then equation (4.4) becomes\vspace{5mm}

$\displaystyle
\left\{    \frac{\Delta(c_{\Lambda})}{2W''(\tilde{c}_{\Lambda})}
           + \frac{2S_{d}}{\u{\bc}\u{\depid}a^{d}d}
                \u{\del}
                  \ln
\frac{\u{\del}+(a\epsilon)^{d/2}}{\u{\del}}
\right.$
$$
\left.     - \frac{\Delta(c_{\Lambda})}{2W''(\tilde{c}_{\Lambda})}
\partial_{c}I_{d}^{(>\epsilon)}(\hat{c}_{\Lambda})
           + E(d,\ \sigma) V^{-(d-\sigma)/d}
\right\}
        V\u{\bc}(\lambda)\u{\del}
           = 1.
\vspace{-5mm}$$\\
Hence
$$
    \u{\del} \simeq \frac{1}{V^{1/2}\ln V}\ \ .
\eqno\rm(4.11)
$$\\
$\displaystyle
        c.\ \ \ \sigma<d<2\sigma.
$
The equation (4.4) has now the form\vspace{5mm}

$\displaystyle
\left\{    \frac{\Delta(c_{\Lambda})}{2W''(\tilde{c}_{\Lambda})}
         + N_{1}(\u{\del})^{d/\sigma-1}+N_{2}\u{\del}
\right.$
$$
\left.   - \frac{\Delta(c_{\Lambda})}{2W''(\tilde{c}_{\Lambda})}
\partial_{c}I_{d}^{(>\epsilon)}(\hat{c}_{\Lambda})
         + E(d,\ \sigma) V^{-(d-\sigma)/d}
\right\}
        V\u{\bc}(\lambda)\u{\del}
           = 1.
\vspace{-5mm}$$\\
Consequently,
$$
    \u{\del} \simeq V^{-d/\sigma}.
\eqno \rm (4.12)
$$\\
Hence one gets the statement of the Lemma. \hspace{55mm}$\Box$
We turn now to fluctuation operators.\vspace{-4mm}\\
\proclaim Proposition 4.1. If $(T,\ \lambda)$ belongs to the critical
line $(T_{c}(\lambda),\ \lambda)$ and $T_{c}(\lambda)>0$, then the
momentum fluctuation operator $F_{0}(P)$ is normal, while the
displacement
fluctuation operator $F_{\delta}(Q)$ is abnormal with a
critical exponent $\delta$ which for $0<\sigma<2$ is dimension and
power-law interactions  dependent:
$$
\delta = \left\{
                \begin{array}{lc}
                 \displaystyle\frac{\sigma}{2d} & \displaystyle
                                               \sigma<d<2\sigma\\
[7mm]
                 \displaystyle \frac{1}{4}+0 &   \displaystyle
                                                   d=2\sigma  . \\
[7mm]
                 \displaystyle \frac{1}{4}  &  \displaystyle
                                                  2\sigma<d
                  \end{array}
         \right.
\eqno{(4.13)}
$$\par \vspace{7mm}
\noindent
{\it Proof}.
For the momentum fluctuation operator, the computation of the variance
 (3.7)
for $T_{c}(\lambda)>0$ yields
$$
\lim_{\Lambda}\eta_{\Lambda}
                  \left(
                      \left\{
                             \frac{1}{V^{\frac{1}{2}+\delta'}}
                               \sum_{l\in\Lambda}(P_{l}-\eta_{\Lambda}
(P_{l}))
                      \right\}^{2}
                  \right)
 = \lim_{\Lambda}  \frac{1}{V^{2\delta'}} m k T_{c}(\lambda).
$$\\
Hence, the variance is finite only if $\displaystyle \delta'=0$. For
the displacement fluctuation operator, the computation of the variance
(3.6) yields
$$
\lim_{\Lambda}\eta_{\Lambda}
                  \left(
                      \left\{
                             \frac{1}{V^{\frac{1}{2}+\delta}}
                               \sum_{l\in\Lambda}(Q_{l}-\eta_{\Lambda}
(Q_{l}))
                      \right\}^{2}
                  \right)
 = \lim_{\Lambda}  \frac{1}{V^{2\delta}} \frac{k T_{c}(\lambda)}
{\u{\del}}.
$$\\
Therefore, the variance is finite for $\displaystyle
\delta=\frac{\gamma}{2}$. By the preceding Lemma one gets immediately
the results annonced in the proposition.$\hspace{4cm}\Box$\\
As far as we get $\delta>0$ and $\delta'=0$ for $T_{c}(\lambda)>0$,
the algebra of fluctuation operators is abelian , see (3.3). It
manifests the classical character of the fluctuations on the part of
the critical line which corresponds $T_{c}(\lambda)>0$ [8].\\
\proclaim Corollary 4.1. (short-range interactions: $\sigma\geq2$)\\
                      Let $(T,\ \lambda)$ belongs to the critical line
$(T_{c}(\lambda),\ \lambda)$ with $T_{c}(\lambda)>0$. The results
for short-range interactions can be deduced from the Proposition 4.1
putting $\sigma=2$, see (3.9) and (4.13).\\
Therefore, the momentum fluctuation operator $F_{0}(P)$ is normal and
the displacement fluctuation operator $F_{\delta}(Q)$ is
critical with a critical exponent
$$
\delta= \left\{
               \begin{array}{lc}
                  \displaystyle \frac{1}{d}   &  2<d<4 \\[7mm]
                  \displaystyle \frac{1}{4}+0 &   d=4 . \\[7mm]
                  \displaystyle \frac{1}{4}   &   4<d
               \end{array}
       \right.
\eqno{(4.14)}
$$\par \vspace{7mm}
\noindent
This means that for the fluctuation operators the inverse power-law
interactions with $\sigma \geq 2$ belong to the same class
universality as the short-range interactions [2,3]. $\hspace{10cm}
\Box$\\
We summarize our results in Fig.2.
\section{\bf Quantum critical fluctuations}

\hspace{6mm}
For the point $T_{c}(\lambda_{c})=0$, we proceed in the same way as in
the study of the rest of the critical line. First we compute the
asymptotic volume behavior of the gap $\u{\del} \simeq V^{-\gamma}$.\\
\proclaim Lemma 5.1. Let $(T,\ \lambda)$ coincides with the point
$(T_{c}(\lambda_{c})=0,\
\lambda_{c})$. Then for $0<\sigma<2$ the asymptotic volume behavior
of
the gap $\Delta(c_{\Lambda}(T_{c}=0,\ h=0))$ is defined by
$$
\gamma=\left\{
               \begin{array}{lc}
                 \displaystyle \frac{\sigma}{d} & \displaystyle
\frac{\sigma}{2}<d<\frac{3\sigma}{2}\\[7mm]
                 \displaystyle \frac{2}{3}+0  & \displaystyle
\frac{3\sigma}{2}=d .\\[7mm]
                 \displaystyle \frac{2}{3}  & \displaystyle
\frac{3\sigma}{2}<d
               \end{array}
       \right.
$$\par \vspace{7mm}
\noindent
{\it Proof}. In the quantum limit $T\rightarrow0$, the
self-consistency
equation (2.3) for $h=0$ becomes
$$
c_{\Lambda}=\frac{\lambda}{2 V \sqrt{\u{\del}}}
            +\frac{1}{V}\sum_{q\neq0}\frac{\lambda}{2\Omega_{q}
(\u{\cl})}.
\eqno{(5.1)}$$\vspace{-2mm}\\
Using the same line of reasoning as in the proof of
Lemma 4.1, we represent (5.1) as
%
  \begin{eqnarray}
            \{  c_{\Lambda}-c^{*} \}
           &+& \{ c^{*}-I_{d}(c_{\Lambda},\ 0,\ \lambda_{c}) \}
\nonumber\\[7mm]
%
          &  + &  \{ I_{d}(c_{\Lambda},\ 0,\ \lambda_{c})
                    -\frac{1}{V}\sum_{q\neq0}\frac{\lambda}
{2\Omega_{q}(c_{\Lambda})}
                  \}     \nonumber \\[7mm]
          &  = &  \frac{\lambda}{2V\sqrt{\u{\del}}},\nonumber
       \end{eqnarray}\\ \vspace{-23mm}
$$\eqno{(5.2)}$$\vspace{-3mm}
where$$\vspace{-3mm}
I_{d}(c_{\Lambda},\ 0,\ \lambda_{c}(0))=
\frac{1}{\u{\depid}}\int_{{\cal B}_{d}}\frac{\lambda}{2 \Omega_{q}
(\u{\cl})}d^dq.
\eqno{(5.3)}$$\\
As in the proof of the Lemma 4.1 one has to take care about
integration around the singular point $q=0$, where $\displaystyle
\Omega_{q}^{2}(\u{\cl})=\u{\del}+a^{\sigma}q^{\sigma}+o(q^{\sigma})$.
The integral (5.3) is finite for
$\u{\del}\rightarrow 0$ if $\displaystyle \frac{\sigma}{2}<d$. Then
we can represent the
second term in the left-hand side of (5.2) as \\
$\displaystyle
\u{\idem}-\u{\idlm}=\frac{\lambda}{2(2\pi)^{d}}
                    \int_{|q|<\epsilon}d^{d}q
                       \{\frac{1}{(aq)^{\sigma/2}}
                        - \frac{1}{\sqrt{(aq)^{\sigma}+\u{\del}}}
                       \}$\\[5mm]
$\displaystyle
=\frac{S_{d}\lambda}{2(2\pi)^{d}}
 \left\{
   \frac{1}{a^{\sigma/2}}\int_{0}^{\epsilon}q^{d-1-\sigma/2}\,dq
  -\frac{1}{a^{\sigma}}\int_{0}^{\epsilon}q^{d-1-\sigma}
         \sqrt{(aq)^{\sigma}+\u{\del}}\,dq
 \right.
$
$$
  \left.
  +\frac{\u{\del}}{a^{\sigma}}\int_{0}^{\epsilon}
          \frac{q^{d-1-\sigma}}{\sqrt{(aq)^{\sigma}+\u{\del}}}\,dq
 \right\}
\vspace{-4mm}$$\\
$
\displaystyle
=\frac{S_{d}\lambda}{2(2\pi)^{d}}
             \{J_{1}-J_{2}+J_{3}\}.
$\vspace{-10mm}
$$\eqno{(5.4)}$$\\
For the first integral we get explicitily:
$$
J_{1}=\frac{1}{a^{\sigma/2}}\int_{0}^{\epsilon}q^{d-1-\sigma/2}\,dq
=\frac{\epsilon^{d-\sigma/2}}{a^{\sigma/2}(d-\sigma/2)}.
\eqno{(5.5)}$$\\
After the change of the variable $\displaystyle
x=\frac{(aq)^{\sigma}}{\Delta}$ we represent the second integral in
the form:
$$
J_{2}
=\frac{\Delta^{d/\sigma-1/2}}{a^{d}\sigma}
   \{
      \int_{0}^{A}x^{d/\sigma-2}\sqrt{1+x}\,dx
     +\int_{A}^{(a\epsilon)^{\sigma}/\Delta}
              x^{d/\sigma-2}\sqrt{1+x}\,dx
    \}, \eqno{(5.6)}
$$
where $0<A<1$.\\
Similary, we get for the third integral:
$$
J_{3}=\frac{\Delta^{d/\sigma-1/2}}{a^{d}\sigma}
   \{
      \int_{0}^{A}\frac{x^{d/\sigma-2}}{\sqrt{1+x}}\,dx
     +\int_{A}^{(a\epsilon)^{\sigma}/\Delta}
                             \frac{x^{d/\sigma-2}}{\sqrt{1+x}}\,dx
    \}.\eqno{(5.7)}
$$\\
In spite of divergencies in (5.6) and (5.7) for $d<\sigma<2d$, the
sum
$J_{1}-J_{2}+J_{3}$ is well-defined for all $0<\sigma<2d$ and has the
following asymptotic for $\u{\del}\rightarrow 0$:
$$
J_{1}-J_{2}+J_{3}= B_{1}(A,\ d,\ \sigma)\Delta^{d/\sigma-1/2}
(\u{\cl})
                  +B_{2}(A,\ d,\ \sigma)\u{\del}.
\vspace{-3mm}
$$\\
Hence, one gets for (5.4) asymptotic:    \vspace{3mm}
$$
\u{\idem}-\u{\idlm}=M_{1}(\u{\del})^{d/\sigma-1/2}+M_{2}\u{\del},
\eqno{(5.8)}
\vspace{-3mm}
$$\\
where $M_{1}=M_{1}(d,\ \sigma)$ and $M_{2}=M_{2}(d,\ \sigma)$.
To analyse equation (5.2) for $0<\sigma<2$ we have to distinguish
three cases (see Fig.3):\\
$a.\ \ \displaystyle \frac{3\sigma}{2}<d$. In the expression (5.8),
only the second term contributes when
$\u{\del}\rightarrow0$:
$$
\u{\idem}-\u{\idlm} \simeq M_{2} \u{\del}.
\eqno{(5.9)}
$$\\
$\displaystyle b.\ \ \frac{3\sigma}{2}=d$. In this case, one can
explicitly calculate the asymptotic of $(J_{1}-J_{2}+J_{3})$ for
$\u{\del}\rightarrow0$ using change of the variable $\displaystyle
x^{2}=\frac{(aq)^{\sigma}}{\Delta}$:\vspace{5mm}\\
$$
\u{\idem}-\u{\idlm} = -\frac{3\lambda S_{d}}{8 \u{\depid} a^{d} d}
                         \u{\del} \ln {\left( \u{\del} \right)}.
\eqno{(5.10)}
$$\\
$\displaystyle c.\ \ \frac{\sigma}{2}<d<\frac{3\sigma}{2}$. The
asymptotic behavior of $\u{\idem}-\u{\idlm}$ is determined by the
first term in the representation (5.8):
$$
\u{\idem}-\u{\idlm} \simeq M_{1} (\u{\del})^{d/\sigma-1/2}.
\eqno{(5.11)}
$$\\
To estimate the last term in the left-hand side of the equation (5.2),
we again refer to [11].\\
For $\displaystyle
\Omega_{q}^{2}(\u{\cl})=\u{\del}+a^{\sigma}q^{\sigma}+o(q^{\sigma})$,
$q\rightarrow0$, the difference between integral $I_{d}$ and sum
vanishes when $V \rightarrow \infty$ as $\displaystyle
V^{-(d-\sigma/2)/d}$:
$$
I_{d}(c_{\Lambda},\ 0,\
\lambda_{c}(0))-\frac{1}{V}\sum_{q\neq0}\frac{\lambda}{2\Omega_{q}
(\u{\cl})}=E(d,\
\sigma/2) V^{-(d-\sigma/2)/d}+o(V^{-(d-\sigma/2)/d}).
\eqno{(5.12)}\vspace{-4mm}$$\\
Substituting the asymptotics (5.9)---(5.12) in the equation (5.2), we
obtain for the volume dependence of the gap at $T_{c}(\lambda_{c})$:
$$
  \u{\del}= \left\{
                  \begin{array}{lc}
  \displaystyle    V^{-\sigma/d}&          \displaystyle
                               \frac{\sigma}{2}<d<\frac{3\sigma}{2}\\
[5mm]
  \displaystyle\frac{1}{V^{2/3}\ln V} &    \displaystyle
                               \frac{3\sigma}{2}=d  .
                                                                   \\
[5mm]
 \displaystyle             V^{-2/3}   & \displaystyle
                                \frac{3\sigma}{2}<d
                  \end{array}
            \right.
$$\\
This gives desires values for the exponent $\gamma$.\hspace{56mm}
$\Box$\\
Now we can  enunciate proposition about the critical exponents
$\delta$ and $\delta'$.\\
\proclaim Proposition 5.1. If $(T,\ \lambda)$ coincides with the point
$(T_{c}(\lambda_{c})=0,\ \lambda_{c})$, then the displacement
fluctuation operator $F_{\delta}(Q)$ is abnormal with a
critical exponent which is for $0<\sigma<2$ dimension and power-law
interactions dependent (see Fig.3):
$$
   \delta=  \left\{
                   \begin{array}{lc}
  \displaystyle   \frac{\sigma}{2d}& \ \ \displaystyle\frac{\sigma}
{2}<d<\frac{3\sigma}{2}\\[5mm]
\displaystyle   \frac{1}{6} +0    &  \ \ \displaystyle
\frac{3\sigma}{2}=d .\\[5mm]
\displaystyle   \frac{1}{6}      & \ \ \displaystyle\frac{3\sigma}
{2}<d
                    \end{array}
             \right.
\eqno{(5.13)}\vspace{5mm}$$\\
Furthermore, the momentum fluctuation operator $F_{\delta'}$ is
supernormal (squeezed) with a critical exponent $\delta'=-\delta$.
\par
\vspace{5mm}\noindent
{\it Proof}. The computation of the displacement fluctuation variance
 (3.6) for
$\beta\rightarrow\infty$ yields
$$
\lim_{\Lambda}\eta_{\Lambda}
                  \left(
                      \left\{
                           \frac{1}{V^{\frac{1}{2}+\delta}}
                           \sum_{l\in\Lambda}(Q_{l}-\eta_{\Lambda}
(Q_{l}))
                      \right\}^{2}
                  \right)
           = \lim_{\Lambda}\frac{1}{V^{2\delta}}
                       \frac{\lambda}{\sqrt{\Delta(c_{\Lambda})}}\ \ .
$$\\
But at the point $(T=0,\ \lambda=\lambda_{c})$, in the thermodynamic
limit $\u{\del}\rightarrow0$ as $V^{-\gamma}$. So, the variance will
be nontrivial only for $\displaystyle\delta=\frac{\gamma}{4}$.
Using the results of the Lemma 5.1, one finds the values
$\delta$ (5.13).\\
For the momentum fluctuation operator, the computation of the
variance~(3.7) for $\beta=\infty$ yields
$$
\lim_{\Lambda}\eta_{\Lambda}
                  \left(
                      \left\{
                          \frac{1}{V^{\frac{1}{2}+\delta'}}
                           \sum_{l\in\Lambda}(P_{l}-\eta_{\Lambda}
(P_{l}))
                      \right\}^{2}
                  \right)
           = \lim_{\Lambda}\frac{1}{V^{2\delta'}}
                     \frac{\lambda m}{2}\sqrt{\Delta(c_{\Lambda})}\
\ .
$$\\
Therefore, the variance will be nontrivial only for $\displaystyle
\delta'=-\delta=-\frac{\gamma}{4} $.\hspace{20mm}$\Box$\\
The nontriviality of the commutation relation is immediate
from (3.3) and $\delta=-\delta'$. So, the algebra of the fluctuation
operators is non-abelian in this part of the critical line.\\
\proclaim Corollary 5.1. (short-range interactions: $\sigma\geq2$)\\
Let $(T,\ \lambda)$ coincide with the point $(T_{c}(\lambda)=0,\
\lambda_{c})$. The results for the short-range interactions can be
deduced from the Proposition 5.1 putting $\sigma=2$, see (3.9) and
(5.13).\\
Therefore, the displacement fluctuation operator $F_{\delta}(Q)$ is
abnormal critical and the momentum fluctuation $F_{\delta'}(P)$ is
supernormal (squeezed) critical and the exponent $\delta=-\delta'$
where (see Fig.3)
$$
  \delta= \left\{
                  \begin{array}{lc}
       \displaystyle              \frac{1}{2d} & 1<d<3\\[7mm]
      \displaystyle               \frac{1}{6}+0 &  3=d .\\[7mm]
   \displaystyle                  \frac{1}{6}   &  3<d
                   \end{array}
           \right.
\eqno{(5.14)}$$\par
\vspace{5mm}\noindent
So, again the inverse power-law interactions with $\sigma\geq2$ give
the same results for the exponents as the short-range ones [2,3], see
Fig.3.\hspace{35mm}$\Box$\\

\section{\bf Concluding remarks}

\hspace{6mm}
For the model (2.1) algebra of the fluctuation operators \linebreak
$\displaystyle F({\cal A})=\{ F_{\delta_{A}}(A)\}_{A\in {\cal A}}$
 (see
Sect.3) is generated by the local canonical operators $Q_{l}$ and
$P_{l}$ where $l,\ l' \in \u{\zens}^{d}$. As far as the state
$\eta(\cdot)$ (see Sect.2) is quasi-free and $\eta(P_{l'})=0,\
\eta(Q_{l}P_{l'})=0$, the algebra $F({\cal A})$ is completely defined
by the variances (3.2) of the fluctuations of the canonical operators
and by their commutators (3.3).

For the model (2.1) with long-range inverse power-law $(0<\sigma<2)$
interactions the algebra $F({\cal A})$ of critical fluctuation
operators is abelian (non-abelian) in the same domain of the critical
line as for the model with short-range interactions (or for
$\sigma\geq2$). On the other hand, the corresponding operator critical
exponents are different and depend on the parameters $\sigma$ and $d$,
 see Fig.2
and Fig.3. In contrast to that, collective excitations in the model
(2.1) are less robust. They depend on the range driven by $\sigma$ in
the both domains of the critical line: $\lambda\rightarrow0$
(''classical limit'') and $T\rightarrow0$ (''quantum limit'')
[12].\vspace{7mm}\\
{\Large \bf Acknowledgements}

One of us (A.C.) is grateful to the Centre de Physique Th\'{e}orique
-Luminy for hospitality extended to him during the D.E.A. stage.
\newpage

\begin{center}
   {\Large \bf Figure Captions}
\end{center}
Figure 1: Phase diagram of the model (2.1).\\[5mm]
Figure 2: Dependence of the displacement fluctuation
operator exponent $\delta$ on $\sigma$ and $d$ on the critical line
for $T_{c}(\lambda)>0$ (classical fluctuations).\\[5mm]
Figure 3: Dependence of the displacement fluctuation operator exponent
$\delta$ on $\sigma$ and $d$ on the critical line $T_{c}(\lambda_{c})
=0$
(quantum fluctuations).

\newpage
\begin{figure}
  \begin{center}
       {\input{fig1}}\\
       {Figure 1}
  \end{center}
\end{figure}

\newpage
\begin{figure}
  \begin{center}
       {\input{fig2}}\\
       {Figure 2}
  \end{center}
\end{figure}

\newpage
\begin{figure}
  \begin{center}
       {\input{fig3}}\\
       {Figure 3}
  \end{center}
\end{figure}

\end{document}